\documentclass[prd,twocolumn,amsmath,amssymb,superscriptaddress,floatfix,nofootinbib]{revtex4}
\usepackage{mathrsfs,bm}
\usepackage{longtable,lscape}
\usepackage{txfonts}
\usepackage{amssymb}
\usepackage{indentfirst}
\usepackage{graphicx,,booktabs}
\usepackage{multirow, overpic}
\usepackage{color}
\usepackage{amssymb}

\usepackage{color}
\usepackage[bookmarks=true,bookmarksopen=false,plainpages=false,breaklinks=true,
   bookmarksnumbered=true,hypertexnames=false,
   filecolor=blue,urlcolor=three,menucolor=three,
   linkcolor=three,citecolor=blueone, colorlinks,
   anchorcolor=blue,runcolor=pink,frenchlinks=red
   pdfstartview=FitH,pdftitle=title,%
   pdfauthor=author]{hyperref}

\allowdisplaybreaks[4]

\definecolor{cover}{rgb}{0.77,0.87,0.88}
\definecolor{blueone}{rgb}{0.1,0.1,.7}
\definecolor{citec}{rgb}{0.14,0.47,0.09}
\definecolor{two}{rgb}{0.0,0.5,0.}
\definecolor{three}{rgb}{.5,.1,0.15}

\begin{document}

\title{The quasi-fission phenomeonon of double charm $T_{cc}^+$ induced by nucleon}

\author{Jun He}
\email{junhe@njnu.edu.cn}
\affiliation{Department of  Physics and Institute of Theoretical Physics, Nanjing Normal University,
Nanjing 210097, China}
\affiliation{Lanzhou Center for Theoretical Physics, Lanzhou University, Lanzhou 730000, China}

\author{Xiang Liu}
\email{xiangliu@lzu.edu.cn (Corresponding author)}
\affiliation{Lanzhou Center for Theoretical Physics, Lanzhou University, Lanzhou 730000, China}
\affiliation{School of Physical Science and Technology, Lanzhou University, Lanzhou 730000, China}
\affiliation{Key Laboratory of Theoretical Physics of Gansu Province, and Frontiers Science Center for Rare Isotopes, Lanzhou University, Lanzhou 730000, China}
\affiliation{Research Center for Hadron and CSR Physics, Lanzhou University and Institute of Modern Physics of CAS, Lanzhou 730000, China}

\date{\today}
\begin{abstract}
In this work, we study the reaction of a nucleon and a doubly charmed state $T_{cc}$.  Under the assumption of the $T_{cc}$ as a molecular state of $D^{*}D$, the reaction of the nucleon and $T_{cc}$ is mediated by exchanges of $\pi$, $\eta$, $\rho$, and $\omega$ meson, which results in  split of  $T_{cc}$ state with two $D$ mesons in  final state. With the help of the effective Lagrangians, the cross section of $p+T^+_{cc}\to p+D^++D^0$ process is calculated, and a very large cross section can be obtained with very small incoming momentum of proton. It decrease rapidly with the increase of the momentum to about 10 mb at momenta of order of  GeV. Such large cross section suggests that induced by a proton the $T_{cc}^+$ state is very easy to decay and transit to two $D$ mesons. In the rest frame of the  $T_{cc}^+$ state, an obvious accumulation of final $D$ meson at small momentum region can be observed  in predicted Dalitz plot, which is due to the molecular state interpretation of $T_{cc}$ state. This novel quasi-fission phenomenon of double charm molecular $T_{cc}^+$ state induced by a proton can be accessible at the forthcoming PANDA experiment. 

\end{abstract}

\maketitle

\section{Introduction}

Recently, the $T_{cc}^+$ state was discovered by the LHCb Collaboration \cite{LHCb:2021vvq,LHCb:2021auc}. As a  very narrow resonance structure, the $T_{cc}^+$ with the significance over $10\sigma$ exists in the $D^0D^0\pi^+$ invariant mass spectrum, which shows that the $T_{cc}^+$ has the minimal quark content $cc\bar{u}\bar{d}$. The LHCb measurement indicates that the $T_{cc}^+$ has mass difference
and width
\begin{eqnarray*}
\delta &=&m_{T_{cc}^+}-(m_{D^{0}}+m_{D^{*+}})=-273\pm61\pm5^{+11}_{-14}\,\,{\rm keV},\\
\Gamma&=&410\pm 165\pm43^{+18}_{-38}\,\, {\rm keV},
\end{eqnarray*}
respectively. The observation of the $T_{cc}^+$ confirmed the existence of double charm tetraquark \cite{Heller:1986bt,Carlson:1987hh,Silvestre-Brac:1993zem,Semay:1994ht,Moinester:1995fk,Pepin:1996id,Gelman:2002wf,Vijande:2003ki,Janc:2004qn,
Navarra:2007yw,Vijande:2007rf,Ebert:2007rn,Lee:2009rt,Yang:2009zzp,Li:2012ss,Luo:2017eub,Karliner:2017qjm,Eichten:2017ffp,Wang:2017uld,Park:2018wjk,
Junnarkar:2018twb,Deng:2018kly,Liu:2019stu,Maiani:2019lpu,Yang:2019itm,Tan:2020ldi,Lu:2020rog,Braaten:2020nwp,Gao:2020ogo,Cheng:2020wxa,Noh:2021lqs,
Faustov:2021hjs}, and inspired further discussions of its properties \cite{Agaev:2022ast,Ke:2021rxd,Chen:2021spf,Deng:2021gnb,Zhao:2021cvg,Santowsky:2021bhy,Albaladejo:2021vln,Chen:2021cfl,Dong:2021bvy,Chen:2021vhg,
Guo:2021yws,Agaev:2021vur,Meng:2021jnw,Fleming:2021wmk,Yan:2021wdl,Ling:2021bir,Hu:2021gdg,Jin:2021cxj,Feijoo:2021ppq,Ren:2021dsi,Chen:2021tnn,Huang:2021urd}.
The main reason why double charm tetraquark attracts the attention from both theorist and experimentalist is that double charm tetraquark is a manifestly exotic state, which can be distinguished from the conventional hadron. The zoo of exotic hadronic state becomes more abundant with adding the reported $T_{cc}^+$.

When facing this novel phenomenon, different proposals to the inner structure of the the $T_{cc}^+$ were proposed. At present, the molecular state \cite{Agaev:2022ast,Ke:2021rxd,Chen:2021spf,Deng:2021gnb,Zhao:2021cvg,Santowsky:2021bhy,Albaladejo:2021vln,Chen:2021cfl,Dong:2021bvy,Chen:2021vhg,
Ren:2021dsi,Chen:2021tnn} and compact
tetraquark \cite{Guo:2021yws,Agaev:2021vur} are two typical assignments to the $T_{cc}^+$, which  are competing with each other. The present experimental data cannot be applied to
distinguish them. Under different assignment to the the $T_{cc}^+$, the investigations of the mass spectrum \cite{Agaev:2022ast,Ke:2021rxd,Chen:2021spf,Deng:2021gnb,Zhao:2021cvg,Santowsky:2021bhy,Albaladejo:2021vln,Chen:2021cfl,Dong:2021bvy,Chen:2021vhg,Guo:2021yws,
Agaev:2021vur,Ren:2021dsi,Chen:2021tnn}, decay behavior \cite{Meng:2021jnw,Fleming:2021wmk,Yan:2021wdl,Ling:2021bir}, and production mechanism \cite{Hu:2021gdg,Jin:2021cxj,Feijoo:2021ppq,Huang:2021urd} can  provide some important aspect of the spectroscopy behavior of
the $T_{cc}^+$. However, it is not the whole aspect of exploring hadronic spectroscopy.

In fact, the reaction of the $T_{cc}^+$ with other hadrons can provide useful information to decode the property of the $T_{cc}^+$.
In this work, we find a novel quasi-fission phenomenon of the molecular $T_{cc}^+$ induced by nucleon, which can be as a unique approach to test the molecular structure of the $T_{cc}^+$.
If the $T_{cc}^+$ is the $DD^*$ molecular state \cite{Li:2012ss}, the $T_{cc}^+$ cannot decay into its hadronic components $D$ and $D^*$ which is kinematically forbidden. However, when a proton interacts with the molecular $T_{cc}^+$, the quasi-fission phenomenon can be happened. In this work, we investigate the quasi-fission phenomenon of the molecular $T_{cc}^+$ induced by a proton.

In realistic calculation, we select a simple but typical reaction $p+T_{cc}^+\to p +D^+ +D^0$ to illustrate how this reaction occurs, by which future experimental exploration of such a reaction is suggested. Among the running and forthcoming  experiments, the PANDA experiment at the Facility for Antiproton and Ion Research (FAIR), which is under construction, has a potential to find out
the reaction of $p+T_{cc}^+\to p +D^+ +D^0$. As peculiar phenomenon of the $T_{cc}^+$ molecular state, its quasi-fission reaction behavior can be applied to test the  
molecular state assignment to the $T_{cc}^+$. 

\begin{figure}[hptb]
\begin{center}
	\scalebox{0.9}{\includegraphics[width=\columnwidth]{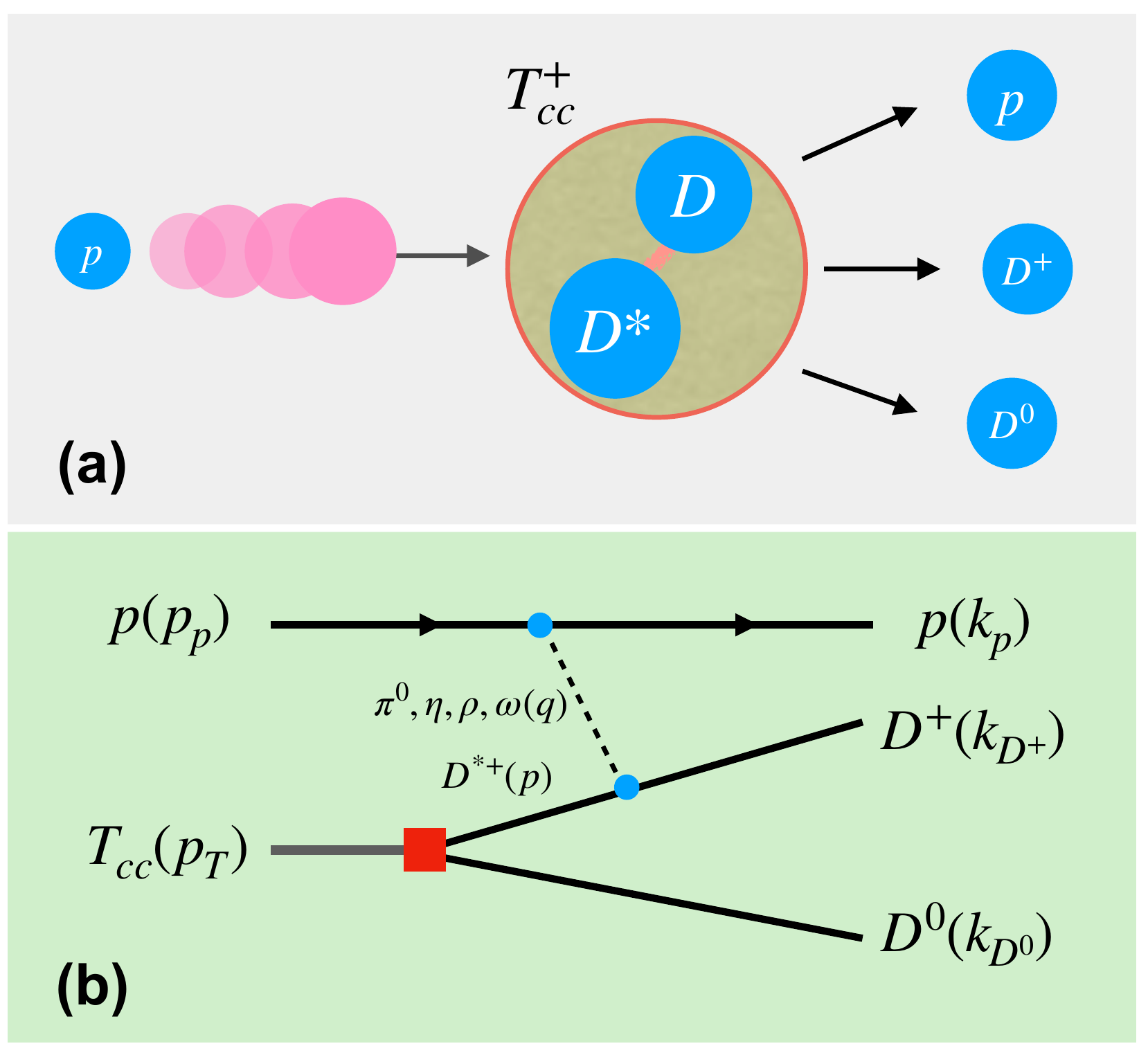}}
	\caption{The reaction of $p+T^+_{cc}\to p+D^++D^0$. a) the sketch map of the reaction. b) the Feynman diagram of the reaction, the denotations of the momenta of  particles are also given.}\label{diagram}
\end{center}
\end{figure}

\section{Reaction of the $T_{cc}^+$ with a proton}

Under the molecular state picture, the $T^+_{cc}$ state can be expressed as
\begin{align}
|T^+_{cc}\rangle&=\frac{1}{\sqrt{2}}\Big(|D^{*+}D^{0}\rangle-|D^{+}{D}^{*0}\rangle\Big).
 \label{Eq: wf}
\end{align}
Attacked by an incoming proton, the $T^+_{cc}$ molecular state can be split into its constituents. The reaction mechanism for the precess $p+T^+_{cc}\to p+D^++D^0$  is illustrated in Fig.~\ref{diagram}. Here, only the first term of the $T^+_{cc}$ molecular  state in the above expression is presented, and the second term can be obtained analogously. The exchanged light mesons are assumed to attack on the vector meson $D^{*+}$, which leads to a  transition to a scalar $D^+$ meson.

In the current work, the most important physical quantity is the cross section of the discussed reaction process, which describes the possibility of the reaction of a proton and a $T_{cc}$ molecular state. In the rest frame of the $T^+_{cc}$,  the cross section for the reaction $p+T^+_{cc}\to p+D^++D^0$ reads as,
\begin{eqnarray}
	d\sigma=\frac{1}{4[(p_1\cdot p_2)^2-m_1^2m_2^2]^{1/2}}\frac{1}{6}\sum_{\lambda_p\lambda_{T_{cc}}\lambda'_p}|{\cal M}_{\lambda_p\lambda_{T_{cc}},\lambda'_p}|^2d\Phi_3,\label{cross}
\end{eqnarray}
with the $p_{1,2}$ and $m_{1,2}$ being the momentum and mass of incoming proton or $T^+_{cc}$. With the help of GENEV code in FAWL, the phase space $d\Phi_3$ is produced  as  $$R_3=(2\pi)^{5}d\Phi_3=\prod^3_i\frac{d^3k_i}{2E_i}\delta^4\left(\sum^n_ik_i-P\right),$$
where the $k_i$ and $E_i$ are the momentum and energy of final particle $i$. The reaction of proton and the $T^+_{cc}$ can be described by a split of the $T^+_{cc}$ into two constituents and the scattering $pD^{*+}\to pD^+$. For the first term of the wave function in Eq. (\ref{Eq: wf}), as shown in Fig.~\ref{diagram} the amplitude of total reaction can be written with an amplitude ${\cal A}_{\lambda_{T_{cc}},\lambda_{{D}^{*+}}}$ for $T_{cc}\to {D}^{*+}{D^0}$ and an amplitude  ${\cal A}'_{\lambda_p\lambda_{\bar{D}^{*+}},\lambda'_p}$ for $p{D}^{*+}\to p{D}^+$ as
\begin{eqnarray}
{\cal M}_{\lambda_p\lambda_{T_{cc}},\lambda'_p}=\sum_{\lambda_{{D}^{*+}}}\frac{{\cal A}_{\lambda_{T_{cc}},\lambda_{{D}^{*+}}}~{\cal A}'_{\lambda_p\lambda_{{D}^{*+}},\lambda'_p}}{p^2-m^2_{D^*}},
\end{eqnarray}
where the $\lambda_p$, $\lambda'_p$, $\lambda_{T_{cc}}$, and $\lambda_{{D}^{*+}}$ are helicities for incoming proton, final proton, initial $T_{cc}$ state, and intermediate ${D}^{*+}$ meson. The $p$ and $m^2_{D^*}$ are the momentum and mass of intermediate ${D}^{*+}$ meson.

First, we need to deduce the amplitude ${\cal A}_{\lambda_{T_{cc}},\lambda_{{D}^{*+}}}$  for the split of the $T^+_{cc}$. The coupling of the molecular state to its constituents can be related to the  binding energy~\cite{Weinberg:1962hj}. Hence, the amplitude can be determined by the scattering length $a$ as~\cite{Braaten:2004fk},
\begin{eqnarray}
{\cal A}_{\lambda_{T_{cc}},\lambda_{{D}^{*+}}}=	\sqrt{\frac{16\pi m_{T_{cc}}m_{D^*}m_D}{\mu^2a}}\epsilon_{\lambda_{T_{cc}}}\cdot\epsilon^*_{\lambda_{{D}^{*+}}},
\end{eqnarray}
where $m_{T_{cc},D^*,D}$ is the mass of the $T_{cc}^+$, and the constituent (${D}^{*+}$ or $D^0$). The $\epsilon_{\lambda_{T_{cc}}}$ and $\epsilon_{\lambda_{{D}^{*+}}}$ are the polarized vectors for $T_{cc}$ and ${D}^{*+}$, respectively.  Scattering length $a=1/\sqrt{2\mu E_B}$ with the reduced mass $\mu=m_{D}m_{D^*}/(m_D+m_{D^*})$ and the $E_B$ being the binding energy. As given in Ref.~\cite{He:2011ed}, the amplitudes  for the $T_{cc}$ splitting with the propagator of $D^*$ meson can be expressed with wave function of the $T_{cc}$, i.e., $\psi({\bm k})=\sqrt{8\pi/a}/({\bm k}^2+1/a^2)$
with normalization $\int d^3k/(2\pi)^3|\psi(k)|^2=1$~\cite{Voloshin:2003nt}, as
\begin{align}
\frac{{\cal A}_{\lambda_{T_{cc}},\lambda_{{D}^{*+}}}}{p^2-m^2_{D^*}}&\simeq-\frac{\sqrt{8m_{T_{cc}}m_{D^*}m_D}}{m_{T_{cc}}-m_D+m_{D^*}}\psi({\bm k}_3)\epsilon_{\lambda_{T_{cc}}}\cdot\epsilon^*_{\lambda_{{D}^{*+}}}.
\end{align}

Besides the split of the $T_{cc}$, we need describe the scattering $pD^{*+}\to pD^+$ as shown in Fig.~\ref{diagram}. To depict the scattering, the following Lagrangians under the heavy quark and chiral symmetries are adopted to construct the vertex of $D^*$, $D$, and light meson~\cite{Casalbuoni:1996pg},
\begin{eqnarray}
 \mathcal{L}_{\mathcal{P}^*\mathcal{P}\mathbb{P}}
&=&- \frac{2g}{f_\pi}(\mathcal{P}^{}_b\mathcal{P}^{*\dag}_{a\lambda}+\mathcal{P}^{*}_{b\lambda}\mathcal{P}^{\dag}_{a})\partial^\lambda{}\mathbb{P}_{ba},\nonumber\\
\mathcal{L}_{\mathcal{P}^*\mathcal{P}\mathbb{V}}
  &=&- 2\sqrt{2}\lambda{}g_V v^\lambda\varepsilon_{\lambda\alpha\beta\mu}
  (\mathcal{P}^{}_b\mathcal{P}^{*\mu\dag}_a +
  \mathcal{P}_b^{*\mu}\mathcal{P}^{\dag}_a)
  \partial^\alpha{}\mathbb{V}^\beta_{ba},
  \end{eqnarray}
 where  ${\mathcal{P}}^{(*)T}
=(D^{(*)+},D^{(*)0})$ is the fields for $D^{(*)}$ meson. And $\mathbb P$ and $\mathbb V$ are two
by two pseudoscalar and vector matrices
\begin{eqnarray}
	{\mathbb P}=\left(\begin{array}{ccc}
		\frac{\sqrt{3}\pi^0+\eta}{\sqrt{6}}&\pi^+\\
		\pi^-&\frac{-\sqrt{3}\pi^0+\eta}{\sqrt{6}}
\end{array}\right),\qquad
\mathbb{V}=\left(\begin{array}{ccc}
\frac{\rho^{0}+\omega}{\sqrt{2}}&\rho^{+}\\
\rho^{-}&\frac{-\rho^{0}+\omega}{\sqrt{2}}
\end{array}\right).\nonumber
\end{eqnarray} The parameters involved here were determined in the literature as $g=0.59$, $\beta=0.9$, $\lambda=0.56$ GeV$^{-1}$, $g_V=5.9$, and $f_\pi=132$~MeV~\cite{Casalbuoni:1996pg,Chen:2019asm}.

 The Lagrangians for the vertex of nucleon and light meson are
 \begin{eqnarray}
  \mathcal{L}_{{\mathbb P}NN}&=&-\frac{g_{ {\mathbb
P}NN}}{\sqrt{2}m_N} \bar{N}_b\gamma_5\gamma_\mu
\partial_\mu{\mathbb P}_{ba}  N_a,\\
\mathcal{L}_{\mathbb{V}NN}&=&-\sqrt{2}g_{\mathbb{V} NN}
\bar{N}_b\bigg(\gamma_\mu+\frac{\kappa}{2m_N}\sigma_{\mu\nu}\partial^\nu\bigg){\mathbb{V}}
_{ba}^\mu N_a,\label{vv}
\end{eqnarray}
where   $N^T=(p,n)$  is field for nucleon.
The coupling
constants $g^2_{\pi NN}/(4\pi)=13.6$, $g^2_{\rho NN}/(4\pi)=0.84$,
$g^2_{\omega NN}/(4\pi)=20$ with
$\kappa=6.1~(0)$ in Eq. (\ref{vv}) for $\rho~(\omega)$ meson, which are
used in the Bonn nucleon-nucleon potential ~\cite{Machleidt:2000ge} and
meson productions in nucleon-nucelon
collision~\cite{Cao:2010km, Tsushima:1998jz,Engel:1996ic}. The $\eta$ exchange is neglected in the current work due to the weak coupling of $\eta$ or
$\phi$ to nucleons as indicated in many previous
works \cite{Machleidt:2000ge,Cao:2010km}.

Applying standard Feynman rules, the amplitude  can be written as
\begin{align}
{\cal M}&= i\frac{\sqrt{8m_Tm_{D^*}m_D}}{m_T-m_D+m_{D^*}}\bar{u}_{\lambda'_p}\tilde{\cal A}_{\lambda_{D^{*+}}}u_{\lambda_p},
\end{align}
with
\begin{align}
\tilde{\cal A}_{\lambda_{D^{*+}}}=&-\frac{1}{2}\left[ P_\rho (q^2)-P_\omega (q^2)\right]\psi({\bm k}_3)i\varepsilon_{\alpha\beta\rho\sigma}v_3^\alpha q^\beta\epsilon^\rho_{\lambda_{D^{*+}}}\gamma^\sigma(1-\rlap\slash q)\nonumber\\
&+\frac{1}{2}\left[P_\rho (q^2)+P_\omega (q^2)\right]\psi({\bm k}_2)i\varepsilon_{\alpha\beta\rho\sigma}v_2^\alpha q^\beta\epsilon^\rho_{\lambda_{D^{*+}}}\gamma^\sigma(1-\rlap\slash q)\nonumber\\
&+\frac{1}{2} \left[\psi({\bm k}_3)\tilde{\epsilon}_T^3\cdot q+\psi({\bm k}_2)\tilde{\epsilon}_T^2\cdot q\right]P_\pi (q^2)\gamma_5\rlap\slash q,
\end{align}
with the abbreviations $\tilde{\epsilon}_T^{3,2}=[\epsilon_T-(k_{2,3}-q)\cdot\epsilon_T~(k_{2,3}-q)/m_{D^*}^2]$ and $v_{3,2}={k_{3,2}}/{\sqrt{m_Dm_{D^*}}}$. Here, the superscripts 2 and 3 are for the first and second parts of the wave function, respectively.
$P_i(q^2)$ is the product of the denominator  of propagator of exchanged  mesons $1/(q^2-m_i^2)$, form factor $f_i(q^2)=(m_i^2-\Lambda^2)/(q^2-\Lambda^2)$, and coupling constant as $F_\pi={\sqrt{2}gg_{ {\mathbb
P}NN}\sqrt{m_Dm_{D^*}}}/{m_Nf_\pi}$ or $F_\mathbb{V}=4g_{\mathbb{V} NN}\lambda g_V\sqrt{m_Dm_{D^*}} $.

\begin{figure}[hptb]
\begin{center}
	\scalebox{0.9}{\includegraphics[width=\columnwidth]{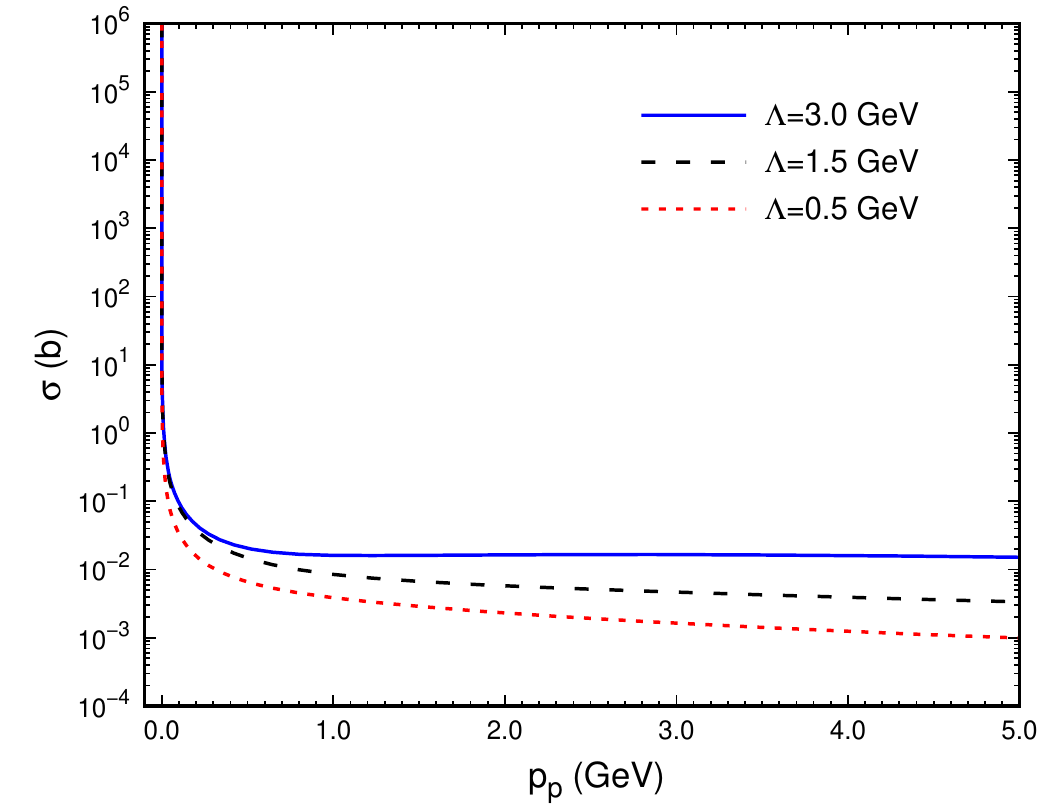}}
\caption{Cross section of the  $p+T^+_{cc}\to p+D^++D^0$ reaction as a function of momentum of incoming proton ${\rm p}_p=|{\bm p}_p|$. The results with cutoffs $\Lambda=3$, 1.5, and 0.5 GeV are given as full (blue), dashed (black), and dotted (red) lines.}\label{pb}
\end{center}
\end{figure}

With the preparation above, we can calculate the cross section of the  $p+T^+_{cc}\to p+D^++D^0$ reaction. The results with cutoffs $\Lambda=0.5$, $1.5$, and 3 GeV are presented in Fig.~\ref{pb}.  One can find that with small momentum of the coming proton, a very large cross section can be obtained. At a momentum of 1 eV, a cross section about $10^7$ b can be reached. Such large cross section is from more reaction time with small speed of incoming proton. With the increase of the incoming momentum, the cross section will decrease very rapidly, and reach a cross section of an order of 10 mb at a momentum about 0.1 GeV. After that, the results becomes relatively stable. In the range of incoming momentum from 1~GeV to 5 GeV, the cross sections with cutoffs $\Lambda=0.5$, $1.5$, and 3 GeV are of an order of magnitude of 1 to 10 mb.

\begin{figure}[hptb]
\begin{center}
	\scalebox{0.97}{\includegraphics[width=\columnwidth]{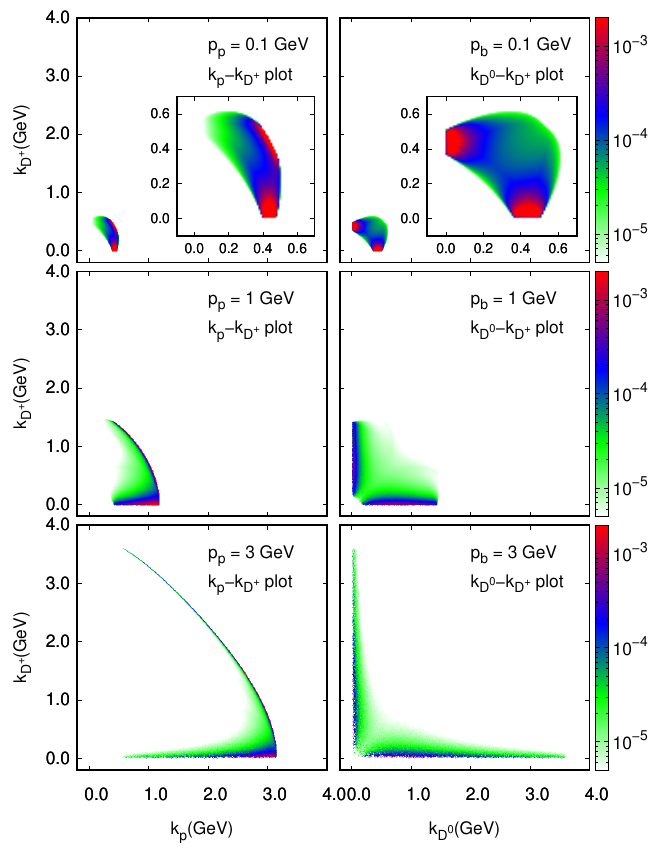}}
\caption{The momentum distributions of final particles for the $p+T^+_{cc}\to p+D^++D^0$ reaction with cutoff $\Lambda=1$ GeV. For each example choice of $\rm p_p$, the figures represent the ${\rm k}_p-{\rm k}_{D^+}$ (left) and ${\rm k}_{D^0}-k_{D^+}$ (right) planes, showing the momentum  ${\rm k}_i=|{\bm k}_i|$  of the final meson $i$. The colorbox means the ratio of event number in a bin of 0.01 GeV$\times$0.01~GeV to the total number of events. The results are obtained with $10^8$ simulation.}
\label{distributions}
\end{center}
\end{figure}

In Fig.~\ref{distributions}, we present the momentum distributions of final particles for the $p+T^+_{cc}\to p+D^++D^0$ reaction at the momentum of incoming proton ${\rm p}_p$=0.1, 1, and 3~GeV.  As expected, the distributions of the  momenta become broader
with the increase of the momentum of incoming proton. At a momentum of 0.1~GeV, the momenta of final particles are in a range smaller than 0.5 GeV while at a large momentum such as 3~GeV, the final particle can have a momentum about 3 GeV.

As show in the left panels in Fig.~\ref{distributions}, the proton after reaction distributes in  a large range of momentum. For example, at ${\rm p}_p$= 3~GeV, the final proton can carry momentum from about 1 to 3~GeV. More events can be observed at high momentum range, which means small energy loss.
Due to the symmetry in the wave function, the distributions of final $D^0$ and $D^+$ is analogously. For the diagram in Fig.~\ref{diagram},  which corresponds to the first term of wave function in Eq.~(\ref{Eq: wf}), the final $D^0$ meson is almost unaffected, which is shown in the ${\rm k}_{D^0}-k_{D^+}$ plane as the vertical stripe with ${\rm k}_{D^0}\sim 0$~GeV . The final $D^+$ meson is from the $D^{*+}$ meson  struck by the proton, and has a broader distribution. The horizontal strip in the ${\rm k}_{D^0}-k_{D^+}$ plane reflects the second term of the wave function where $D^{*0}$ is struck and $D^+$ is almost unaffected.

\section{Discussion and conclusion}

As good candidate of exotic states, the newly observed $T_{cc}^+$ \cite{LHCb:2021vvq,LHCb:2021auc}
 not only confirms the former prediction of double charm tetraquark \cite{Heller:1986bt,Carlson:1987hh,Silvestre-Brac:1993zem,Semay:1994ht,Moinester:1995fk,Pepin:1996id,Gelman:2002wf,Vijande:2003ki,Janc:2004qn,
Navarra:2007yw,Vijande:2007rf,Ebert:2007rn,Lee:2009rt,Yang:2009zzp,Li:2012ss,Luo:2017eub,Karliner:2017qjm,Eichten:2017ffp,Wang:2017uld,Park:2018wjk,
Junnarkar:2018twb,Deng:2018kly,Liu:2019stu,Maiani:2019lpu,Yang:2019itm,Tan:2020ldi,Lu:2020rog,Braaten:2020nwp,Gao:2020ogo,Cheng:2020wxa,Noh:2021lqs,
Faustov:2021hjs}, but also  has aroused
theorists' interest in further revealing its property combing experimental data \cite{Agaev:2022ast,Ke:2021rxd,Chen:2021spf,Deng:2021gnb,Zhao:2021cvg,Santowsky:2021bhy,Albaladejo:2021vln,Chen:2021cfl,Dong:2021bvy,Chen:2021vhg,
Guo:2021yws,Agaev:2021vur,Meng:2021jnw,Fleming:2021wmk,Yan:2021wdl,Ling:2021bir,Hu:2021gdg,Jin:2021cxj,Feijoo:2021ppq,Ren:2021dsi,Chen:2021tnn,Huang:2021urd}. Since experimental precision is not enough to definitely conclude whether or not the $T_{cc}^+$ is a $DD^*$ molecular state, we should pay more effort to find peculiar phenomenon relevant to the $T_{cc}^+$ molecular state. Although mass, decay, and production are important aspects to reflect the inner structure of the $T_{cc}^+$, it is not the whole aspect of exploring the $T_{cc}^+$ property. 
Just considering the situation of the study of the $T_{cc}^+$, we propose that the reaction of the $T_{cc}^+$ with a nucleon is an approach to reveal the nature of the $T_{cc}^+$. Focusing on such a research issue, the concete study is still not enough. 

With the great interest of the reaction of the $T_{cc}^+$, in this work, we study the reaction of a nucleon and a double charm $T_{cc}$. Under the assumption of the $T_{cc}$ as a molecular state of $D^{*}D$, the reaction of the nucleon and $T_{cc}$ is mediated by exchanges of $\pi$, $\eta$, $\rho$, and $\omega$ meson, which results in split of  the $T_{cc}^+$ state with two $D$ mesons in final state. With the help of the effective Lagrangians, the cross section of $p+T^+_{cc}\to p+D^++D^0$ process is calculated, and a very large cross section can be obtained with very small incoming momentum of proton. It decrease rapidly with the increase of the momentum to about 10 mb at momenta of order of GeV. Such large cross section suggests that the $T_{cc}$ molecular state is very easy to decay and transit to two $D$ mesons induced by a proton. We call this peculiar phenomenon as quasi-fission of double charm $T_{cc}^+$ molecular state induced by nucleon, which can be applied to test the molecular state assignment to the $T_{cc}^+$. 

In summary, we predicted a quasi-fission phenomenon of double charm $T_{cc}^+$ molecular state induced by nucleon, which can meet the physics aim of the forthcoming PANDA experiment at FAIR.  
Searching for the quasi-fission phenomenon of the $T_{cc}^+$ induced by nucleon can shed light on the nature of the 
$T_{cc}^+$, which is crucial step when constructing exotic hadron family. 

\section*{Acknowledgements}

This work is supported by the China National Funds for Distinguished Young Scientists under Grant No. 11825503, the National Key Research and Development Program of China under Contract No. 2020YFA0406400, the 111 Project under Grant No. B20063, the National Natural Science Foundation of China under Grant No. 12047501, No. 12175091, No. 11965016, No. 11775050 and No. 11775050, CAS Interdisciplinary Innovation Team, and the Fundamental Research Funds for the Central Universities under Grants No. lzujbky-2021-sp24.

\end{document}